# High efficiency event-counting thermal neutron imaging using Gd doped micro channel plate[*]

TIAN Yang(田阳)[1, 2;1)]   YANG Yi-Gang(杨祎罡)[1,2]   PAN Jing-Sheng(潘京生)[3]   LI Yu-Lan(李玉兰)[1,2]
LI Yuan-Jing(李元景)[1,2]

1 (Department of Engineering Physics, Tsinghua University, Beijing 100084, China)
2 (Key Laboratory of Particle & Radiation Imaging (Tsinghua University), Ministry of Education, Beijing 100084, China)
3 (Nanjing Branch, North Night Vision Tech. Co., Ltd., Nanjing 211102, China)

**Abstract:** An event-counting thermal neutron imaging detector based on 3 mol % $^{nat}Gd_2O_3$ doped micro channel plate (MCP) has been developed and tested. Thermal neutron imaging experiment was carried out with a low flux neutron beam. Detection efficiency of 33 % was achieved with only one doped MCP. The spatial resolution of 72 μm RMS is currently limited by the readout anode. A detector with larger area and improved readout method is now being developed.

## 1 Introduction

The concept of direct conversion and detection of thermal neutrons using doped micro channel plate (MCP) was first proposed by Fraser and Pearson in 1990 [1]. The first experimental imaging result was achieved using boron and gadolinium doped MCP by researchers from University of California at Berkeley and Nova Scientific Inc. in 2007 [2]. High spatial and temporal resolution as well as relatively high detection efficiency is then confirmed by a serial of experiments [3-5]. As a result, event-counting detector based on doped MCP combined with pixel readout chip is successfully applied to various neutron imaging experiments [6, 7].

Similar cooperating research is carried out in China by Tsinghua University and NNVT Co., Ltd. Both doping and coating have been attempted to increase the neutron sensitivity of MCP [8]. Our coated MCPs are still confronted with the problem of electron emitting layer integration. The first thermal neutron imaging result of our doped MCP is presented in this paper.

## 2 Neutron imaging detector using wedge strip anode

Although boron have several advantages over gadolinium in neutron conversion, such as shorter range of the charged particles (higher spatial resolution achievable), large cross section over the whole thermal region and simple prompt gamma spectrum (easy for coincidence measurement), highly enriched $^{10}B$ (50000 \$/kg) is needed to realize desirable detection efficiency with practical thickness (around 1 mm) of the MCP. Doping with $^{nat}Gd$ is a cost-effective choice. In our first attempt, 3 mol % $^{nat}Gd_2O_3$ doped MCP is made, which has 50 mm diameter and 0.6 mm thickness. The diameter and pitch of the channel are 10 and 12.5 μm respectively with a porosity ratio of 63 %. Detailed composition and production procedures are included in [9].

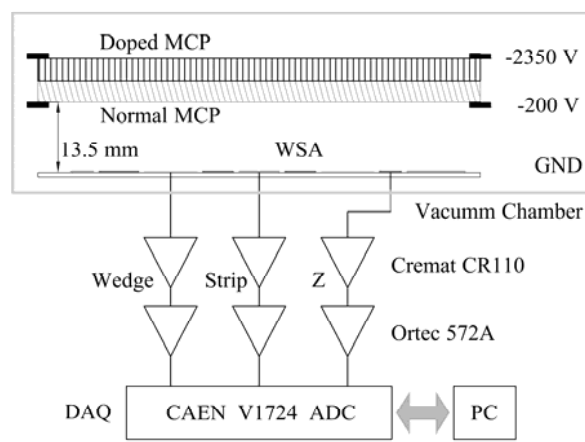

Fig. 1.  The imaging detector and the electronic system.

The $^{nat}Gd$ doped MCP (0 bias angle) is coupled to a normal 0.6 mm MCP (8° bias angle) in the form of a chevron stack. A wedge strip anode (WSA) [10] is employed as a 2-D event-counting readout. These are the essential parts of our neutron imaging detector as illustrated in Fig. 1.The simple electronic system of WSA is comprised of only 3 channels.

---

[*] Supported by the National Natural Science Foundation of China (10735020) and (11075026)
1) Email: cycjty@126.com

## 3 Imaging with low flux neutron beam

Thermal neutron imaging was carried out at the 49-2 reactor at China Institute of Atomic Energy (CIAE). As the reactor is intended for isotope production and sample irradiation, there is no gamma filter in the neutron beam. The n/γ ratio is still as low as $1.07 \times 10^{10}$ n/(cm$^2$•Sv) after 10 cm thick lead brick. This value is one order lower than those of some neutron sources optimized for imaging [11,12]. The thermal neutron flux within the imaging plane is just 123 n/(cm$^2$•s) at a L/D of 360:1 according to the measurement by a $^3$He tube.

Fortunately the dark count rate of our detector is just 0.11 s$^{-1}$•cm$^{-2}$. We managed to obtain a transmission imaging of some high contrast objects (Fig. 2) after irradiating the detector for 2 hours and 39 minutes.

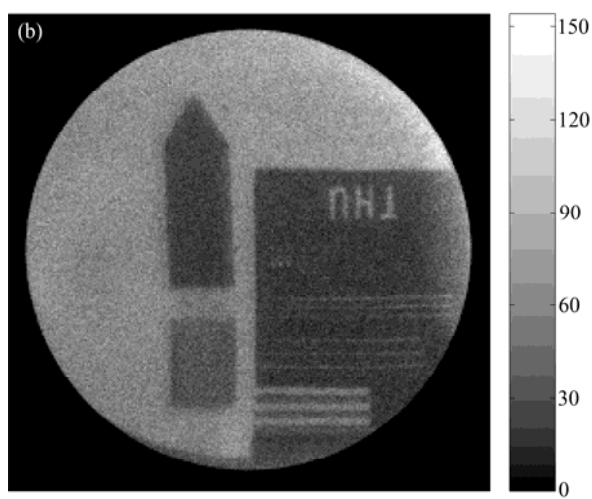

Fig. 2. Imaging objects and results. The imaging objects are fixed in the vacuum chamber. As the neutron flux is very low, in order to get imaging of desirable quality within acceptable acquisition time we select objects of high contrast for thermal neutrons: boron nitride grinding head and 0.5 mm Gd mask in (a). In (b), the pixel size is 100 μm.

## 4 Discussion and conclusion

In Fig.2, after removing the part of the boron nitride grinding head one can get a spectrum of the event density as shown in Fig. 3. There are two well separated peaks corresponding to two different regions: the 0.5 mm Gd blocked part and the unblocked part. Although the 0.5 mm Gd cannot effectively absorb epithermal neutrons, the absorption by the Gd doped MCP is also negligible. Thus in the blocked part it can be considered that only gamma signals and dark counts contribute to the event density. The difference between the two peaks is the net thermal neutron events per pixel in the open area. After comparing this value with the thermal neutron flux (123 n/(cm$^2$•s)), we estimated the detection efficiency to be 33 %. The spatial resolution was derived by fitting the edge (Fig. 4). Resolution of 72 μm RMS was achieved in the central area.

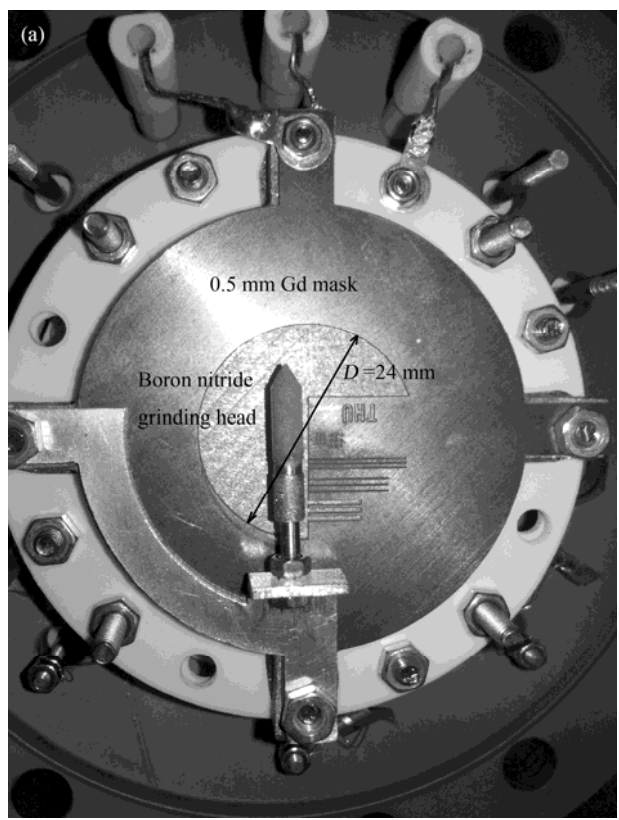

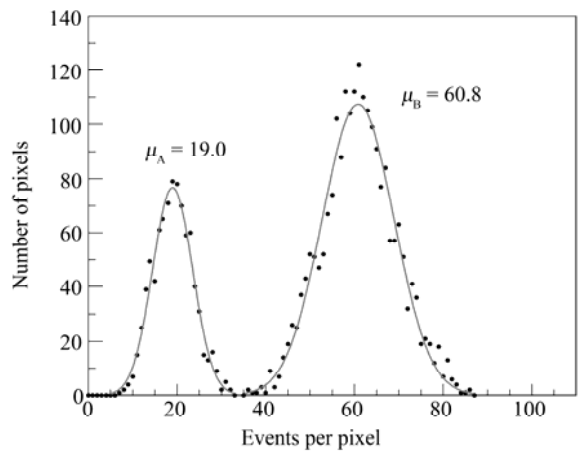

Fig. 3. The modified grey scale spectrum. This spectrum stands for the event density distribution after removing the grinding head. $\mu_A$ and $\mu_B$ are the fitted average events per pixel in the blocked and unblocked parts respectively.

The efficiency is higher than the measured result of the Gd doped MCP by NOVA [13]. Although the value of 33 % is still much lower than the model-predicted 50 % for $^{10}$B doped MCP [14], the efficiency can be further enhanced by increasing the thickness or using smaller channel diameter or both. According to our theoretical model, the efficiency for 25.3 meV neutrons can be increased to 62 % (5 μm channel diameter and 1 mm thickness) [15].

The spatial resolution is mainly limited by the WSA. In our previous tests, the spatial resolution for 10 kVp X-ray remained the same level. The intrinsic event centroiding uncertainty is due to the electron transport process in the MCP. According to the Monte Carlo simulation of such process, the uncertainty is 13 μm RMS for 71.9 keV electrons. Large area Gd doped MCP (10 cm diameter) is now being developed. A 2-D delay line system will be used as its readout. Both the spatial resolution and the imaging area can be greatly improved.

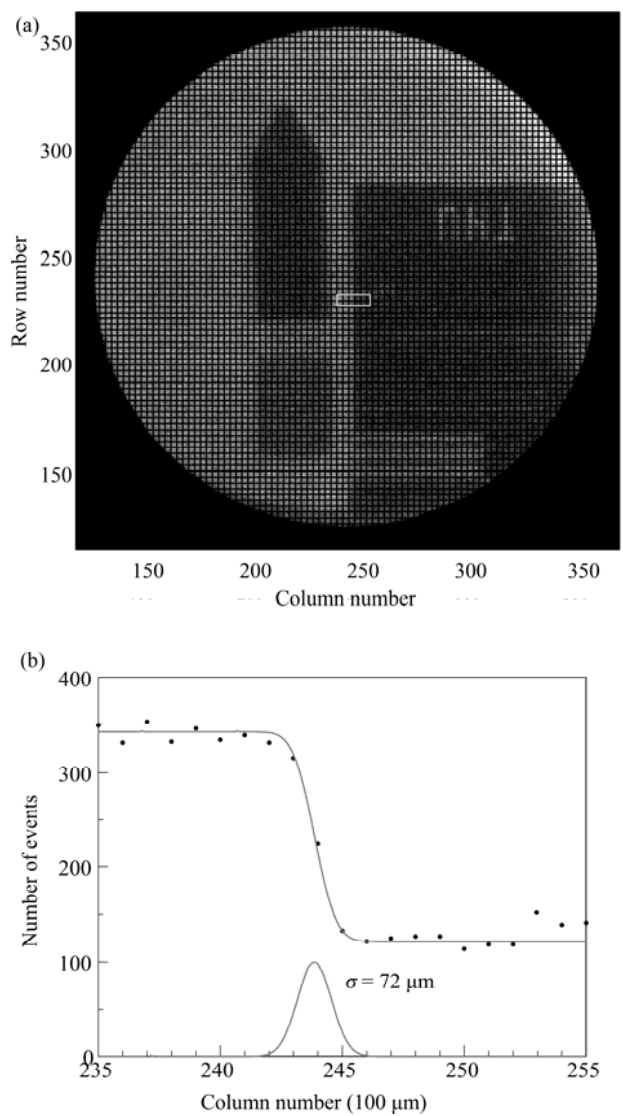

Fig. 4. The derived spatial resolution. In (a), the length of each grid is 3 pixels. The edge within the white box was fitted using error function. The number of events in (b) is the sum of 6 adjacent rows within one column. This was done to suppress the statistics fluctuation at the expense of spatial resolution loss. The RMS of the derivative (Gaussian function) is 72 μm.

The imaging result also testifies that the event-counting detector based on doped MCP is able to work with a low flux and high gamma content neutron beam. It is useful for in situ thermal neutron imaging where only mobile sources of relatively low performance are available.

In conclusion the first event-counting imaging test of our Gd doped MCP detector was successful. Detection efficiency as high as 33 % for thermal neutrons was achieved using only one single 0.6 mm thick doped MCP. The spatial resolution of 72 μm RMS is limited by the WSA and can be greatly enhanced.

________________________________________________

is limited by the WSA and can be greatly enhanced.

________________________________________________

## 基于掺钆微通道板的高探测效率热中子成像研究

TIAN Yang(田阳)[1, 2;]  YANG Yi-Gang(杨祎罡)[1,2]  PAN Jing-Sheng(潘京生)[3]  LI Yu-Lan(李玉兰)[1,2]  LI Yuan-Jing(李元景)[1,2]

1 (清华大学工程物理系，北京 100084，中国)
2 (粒子与辐射成像教育部重点实验室（清华大学），北京 100084，中国)
3 (北方夜视南京分公司，南京 100084，中国)

**中文摘要**：本文基于掺杂 3 mol %天然氧化钆的微通道板制作并测试了一种计数式热中子成像探测器。在低通量的热中子束流上进行的成像实验表明，采用一片上述微通道板其热中子探测效率可达 33 %。受读出电极限制，探测器可实现 72 微米的位置分辨。具有更大面积并采用更先进读出方法的探测器正在研发中。

**关键词**：掺杂微通道板，热中子成像，计数式

**PACS**: 28.20.Pr, 29.40.Gx